# Even-integer Quantum Hall Effect in an Oxide Caused by Hidden Rashba Effect

Jingyue Wang[1†], Junwei Huang[2†], Daniel Kaplan[3,4†], Xuehan Zhou[1†], Congwei Tan[1†], Jing Zhang[5], Gangjian Jin[5], Xuzhong Cong[1], Yongchao Zhu[1], Xiaoyin Gao[1], Yan Liang[1], Huakun Zuo[5], Zengwei Zhu[5], Ruixue Zhu[6], Ady Stern[3], Hongtao Liu[1], Peng Gao[6], Binghai Yan[3*], Hongtao Yuan[2*], Hailin Peng[1*]

[1] Center for Nanochemistry, Beijing Science and Engineering Center for Nanocarbons, Beijing National Laboratory for Molecular Sciences, College of Chemistry and Molecular Engineering, Peking University, 100871, Beijing, China.

[2] National Laboratory of Solid State Microstructures, Collaborative Innovation Center of Advanced Microstructures, College of Engineering and Applied Sciences, and Jiangsu Key Laboratory of Artificial Functional Materials, Nanjing University, 210000, Nanjing, China.

[3] Department of Condensed Matter Physics, Weizmann Institute of Science, 7610001, Rehovot, Israel.

[4] Center for Materials Theory, Department of Physics and Astronomy, Rutgers University, Piscataway, NJ 08854, USA

[5] Wuhan National High Magnetic Field Center and School of Physics, Huazhong University of Science and Technology, 430074, Wuhan, China.

[6] International Center for Quantum Materials and Electron Microscopy Laboratory, School of Physics, Peking University, 100871, Beijing, China.

† These authors contributed equally to this work.

*Corresponding author. E-mail: hlpeng@pku.edu.cn, htyuan@nju.edu.cn, binghai.yan@weizmann.ac.il

**Abstract:**

In the presence of high magnetic field, quantum Hall systems usually host both even- and odd-integer quantized states because of lifted band degeneracies. Selective control of these quantized states is challenging but essential to understand the exotic ground states and manipulate the spin textures. Here, we study the quantum Hall effect in $Bi_2O_2Se$ thin films. In magnetic fields as high as 50 T, we observe only even-integer quantum Hall states, but no sign of odd-integer states. However, when reducing the thickness of the epitaxial $Bi_2O_2Se$ film to one unit cell, we observe both odd- and even-integer states in this Janus (asymmetric) film grown on $SrTiO_3$. By means of a Rashba bilayer model based on *ab* initio band structures of $Bi_2O_2Se$ thin films, we can ascribe the absence of odd-integer states in thicker films to the hidden Rasbha effect, where the local inversion symmetry breaking in two sectors of the $[Bi_2O_2]^{2+}$ layer yields opposite Rashba spin polarizations, which compensate with each other. In the one unit cell $Bi_2O_2Se$ film grown on $SrTiO_3$, the asymmetry introduced by top surface and bottom interface induces a net polar field. The resulting global Rashba effect lifts the band degeneracies present in the symmetric case of thicker films.

**Main Text:**

The quantum Hall effect (QHE) is of great significance for probing and understanding novel electron states in two-dimensional (2D) systems [1-14]. In the integer QHE, the Hall resistance is quantized into integer plateaus $h/ve^2$, where $h$ is the Planck's constant, $e$ is the electron's charge, and $v$ is an integer. In most 2D electronic systems, the coexistence of even- and odd-integer quantum Hall plateaus can be universally observed. [2,5,7,9-13,15-19] Since when the magnetic field $B \geq 30$ T, the Zeeman splitting energy reaches 3.5 meV (if the $g$ factor around 2), which is much larger than thermal (0.36 meV for 4.2 K) and Dingle (0.86 meV for Dingle temperature of 10 K) broadening energies. The similar QHE in these systems indicate the same spin textures with lifted spin degeneracies. Recently, the emergence of 2D layered materials has provided opportunities to explore unique spin textures and QHE, when the electron states and spin textures can be tuned by external modulation of electric field and by the inversion-symmetry engineering at the atomic level [3,4,7,9,13-16,18-21]. However, in the previously reported QHE in 2D layered materials, the quantization in a high magnetic field is mainly prevailed by the Zeeman effect, in which the spin-orbital coupling (SOC) is relatively small due to the light atomic weight therein [14]. In contrast, for a quantum Hall system with stronger SOC and Rashba effect, it is feasible to achieve new spin textures and tailor the QHE when the SOC dominates the electron states.

In this work, we demonstrate the absence of odd-integer state when the magnetic field is up to 50 T in $Bi_2O_2Se$ quantum Hall system driven by a unique spin-degenerated Rashba bilayer structure, *i.e.*, a hidden Rashba effect[22], in which two sectors form inversion partner with opposite Rashba spin polarizations and compensate with each other. We note that the hidden Rashba effect may coexist with the global Rashba effect in $Bi_2O_2Se$ films because of symmetry breaking induced by the substrate/film interface. Both odd- and even-integer quantum Hall plateaus appear when the thickness of epitaxial films with asymmetric Janus structures reduces to 1 unit cell (uc, 1-uc-thick $Bi_2O_2Se$ contains 2 layers), where interface-induced inversion-symmetry breaking generates a giant global Rashba effect. In addition, the global Rashba parameter $\alpha_R^0$ in this 2D Janus structure reaches 440 meV·Å, which is one of the largest values among the known 2D semiconducting Rashba systems. Thus, the compensation status of two Rashba layers and QHE can be effectively tailored by modulating the thickness of high-mobility 2D $Bi_2O_2Se$ films, enabling the control of novel electron states, band topologies and spin textures.

**Crystal structure and hidden Rashba effect in layered $Bi_2O_2Se$**

We choose 2D $Bi_2O_2Se$ as the target material because of the following unique characteristics. Firstly, the electron states near the conduction band minimum originate mainly from the heavy Bi *p*-orbital bands with strong SOC. Thus, 2D $Bi_2O_2Se$ can serve as an ideal platform to study and tailor SOC and spin textures. Secondly, different from van der Waals materials, bulk $Bi_2O_2Se$ belongs to *I*4/*mmm* space group and shows an inversion-symmetric layered lattice structure with tetragonal $[Bi_2O_2]^{2+}$ layers and $[Se]^{2-}$ layers alternately stacked along the *c* axis, forming unique inter-layer dipoles [23-25] (Fig. 1a). For the $[Bi_2O_2]^{2+}$ layer consisting of one O layer sandwiched by two Bi monolayers with the dipole and strong atomic SOC, each Bi monolayer feels a strong Rashba effect while neighboring Bi monolayers exhibit opposite signs of Rashba SOC. Consequently, each $[Bi_2O_2]^{2+}$ layer forms a Rashba bilayer (Fig. 1b), in which the bands are spin-degenerated and the total Rashba spin texture is hidden in the presence of inversion symmetry [22]. At this condition, the Rashba fields of two layers are compensated. Different from the global Rashba effect [14,26-31], the hidden Rashba effect has not been studied in the quantum Hall regime before and will result in unique QHE due to the coupling of opposite SOC layers and the compensated Rashba fields (Fig. 1c). Thirdly, the hidden Rashba effect can be regulated via thickness control, and high-mobility $Bi_2O_2Se$ films can be controllably grown by the molecular beam epitaxy (MBE) technique (see Materials and Methods) with different thicknesses even down to 1 uc (Fig. S1) [32,33]. As shown in Fig. 2a-c, the interfacial $Bi_2O_2Se$ layer in films with different thicknesses is inversion-asymmetric and forms a Janus structure due to the giant dielectric-constant difference and absence of bottom Se at the interface. In the film, electronic wave functions distribute mainly in the middle region (see Fig. S25). Thus, the conduction band is much less affected by the interfacial field in thick films than in ultra-thin films. Then, thick films show negligible global Rashba effect and carry hidden spin texture. As the film thickness decreases, the wavefunction reaches the interface and starts feeling the interfacial electric field. So ultra-thin films (e.g., 1 uc thick) may exhibit a strong global Rashba effect. Consequently, films with different thickness will have different Rashba SOC. Thus, the high-mobility epitaxial 2D $Bi_2O_2Se$ films with tunable thicknesses enable the selective control of the Rashba effect, spin textures and quantum Hall states.

To demonstrate the excellent mobility property in epitaxial $Bi_2O_2Se$ films, we performed low-temperature Hall effect measurements (Fig. 2d), based on which the Hall mobility can be obtained. Three significant observations need to be addressed here. Firstly, the Hall mobility of epitaxial $Bi_2O_2Se$ films at lowest temperature reaches a maximum value of 12,435 $cm^2 \cdot V^{-1} \cdot s^{-1}$, which enables us to observe Shubnikov–de Haas (SdH) oscillations and quantized states therein. Secondly, the Hall mobility increases with increasing thickness. The decrease of carrier mobility in the ultrathin $Bi_2O_2Se$ film is presumably due to the scatterings from the interface and the top surface. Thirdly, the mobility in non-encapsulated $Bi_2O_2Se$ films is higher than that in other reported non-encapsulated 2D semiconductors [13,17,34-36] due to the small effective mass, the excellent chemical stability in ambient conditions and the high-quality crystalline order of the epitaxial $Bi_2O_2Se$ films (Table. S1). The nature of ultrahigh mobility makes 2D $Bi_2O_2Se$ an excellent material platform to study SdH oscillations and QHE.

**Thickness-dependent SdH oscillations and QHE in 2D $Bi_2O_2Se$**

The angular-dependent magnetoresistance measurements were performed to confirm the 2D nature of the epitaxial films. As shown in Fig. 3a, we observed pronounced SdH oscillations in the longitudinal magnetoresistance $R_{xx}$ in the 6-uc-thick epitaxial $Bi_2O_2Se$ film when applying static magnetic fields ($B$) up to 14 T. The positions of the SdH oscillation extrema as a function of the perpendicular component of the field barely move when tilting the angle ($\theta$) between the magnetic field and the normal direction of the sample plane, indicating a 2D feature of the 6-uc-thick $Bi_2O_2Se$ (Fig. 3a) with a single frequency (Fig. S2a). Meanwhile, the angle-dependent oscillation frequency $f$ extracted from the fast Fourier transform (FFT) analysis can be well fitted by a 2D model ($f \propto 1/\cos\theta$), also revealing a strict 2D Fermi surface in the $Bi_2O_2Se$ film (Fig. S2b). Note that the absence of Zeeman splitting induced Landau level (LL) crossing within the experimental maximum $B$ of 14 T indicates a small effective Landé $g$-factor in the 6-uc-thick device [7,14].

Figure 3b shows the quantized states of the Hall resistance ($R_{xy}$) in the 6-uc-thick epitaxial $Bi_2O_2Se$ at 4.2 K in pulsed high magnetic field up to 50 T [37,38]. Remarkably, only even-integer quantum Hall plateaus ($v = 2, 4, 6…$) were clearly observed when tuning the carrier densities through the $SrTiO_3$ (STO) dielectric substrate (Fig. S3 and S4). Similar even-integer QHE was observed in the static magnetic fields at 1.5 K (Fig. S5). To confirm the intrinsic and robust nature of the missing odd-integer quantum Hall plateaus in 2D $Bi_2O_2Se$ systems, free-standing nanoflake of $Bi_2O_2Se$ [39] with thickness up to 9 uc were grown via chemical vapor deposition (CVD) and transferred as van der Waals stacked 2D Hall devices for the electronical measurements. The atomically-flat and high-$\kappa$ 2D layered $Bi_2SeO_5$ dielectric notably improved the mobility of 2D $Bi_2O_2Se$ nanoflake [40]. As shown in Fig. 4a and 4b, only even-integer quantum Hall plateaus ($v = 2, 4, 6, 8, 10…$) in inversion-symmetric $Bi_2O_2Se$ nanoflake clearly emerge down to the quantum limit with filling factor $v = 2$ as the magnetic field increases. This is similar to that the observations in the 6-uc-thick epitaxial film.

We then studied the thickness-dependent evolutions of SdH oscillations and QHE in epitaxial $Bi_2O_2Se$ films. As shown in Fig. 5a and 5b, the $R_{xx}$–$B$ and $R_{xy}$–$B$ measurements of the 6-uc-thick and 2.5-uc-thick $Bi_2O_2Se$ films under pulsed magnetic fields up to 50 and 55 T were performed, respectively. Note that only even-integer quantum Hall plateaus, i.e., $v =2, 4, 6, 8, 10, 12…$, can be observed (Fig. 5a, 5b, Fig. S6a and Fig. S7a). Meanwhile, the consistent QHE in 6- and 2.5-uc-thick films excludes the existence of parallel, independent QHE layers, which is a trivial explanation of the missing of odd-integer quantum Hall states. Otherwise, the plateau step would be proportional to the film thickness. In the 6-uc-thick device, $R_{xx}$ exhibits a strict zero plateau when the magnetic field exceed 42 T, revealing a strict QHE therein. In the 2.5-uc-thick device, $R_{xx}$ almost reaches zero when magnetic field $B = 50$ T. When the film thickness was further reduced to 1 uc, significant SdH oscillations and quantized plateaus can still be observed. Surprisingly, both odd- and even-integer quantum Hall plateaus appeared in the 1-uc-thick device. As shown in Fig. 5c, an odd-integer quantum Hall plateau of $v = 5$ appeared at $B > 40$ T, which is totally different from the absence of odd-integer QHE in thicker films. Meanwhile, two valleys ($v = 6$ and $v = 7$) between the $v = 5$ and $v = 8$ plateaus in $R_{xx}$–$B$ curves can be observed clearly, indicating the fully separation of spin polarizations under high magnetic fields. The coexistence of odd- and even- integer quantum Hall states can also be observed when tuning the carrier density in this device (Fig. S6b and Fig. S7b).

**Analysis of quantum oscillations in 2D $Bi_2O_2Se$ films**

Quantum oscillations can identify the fermiology and resolve the Rashba splitting. For the lowest conduction band of $Bi_2O_2Se$ film, SdH oscillations exhibit a single frequency for the ideal hidden Rashba effect but reveal a sizable frequency splitting ($\Delta f$) for the global Rashba effect. We firstly performed FFT analysis of quantum oscillations in $R_{xx}$ to get the information about the 2D Fermi surfaces (Fig. 5d). For the 6-uc-thick device, the quantum oscillations exhibit single frequency ($f$) as shown in top panel in Fig. 5d, revealing negligible spin splitting and spin-degenerated band structure. For the 2.5-uc-thick device, the quantum oscillations show minor splitting ($\Delta f = 11$ T) in the FFT spectrum. The splitting behavior may arise from the formation of inner-outer Fermi surfaces and indicates a weak global Rashba effect. In contrast, for the 1-uc-thick device, the quantum oscillations show a distinct beating behavior of $\Delta f > 60$ T, suggesting two distinct bands therein.

Landau fan diagrams were further performed to confirm the spin degeneracies in epitaxial $Bi_2O_2Se$ films. The slopes ($k$) in the Landau fan diagram correspond to the carrier densities of bands ($n_{2D} = s \times k \times e/h$, where $n_{2D}$ is the carrier density, $s$ is the spin degeneracy). In the 6-uc-thick film, the slope in Landau fan diagram is 49.4 T (Fig. 5e), which correspond to carrier densities of $s \times 1.2 \times 10^{12}$ $cm^{-2}$, respectively. Meanwhile, the carrier densities obtained from the Hall curves are $2.6 \times 10^{12}$ $cm^{-2}$. Thus, by comparing the carrier densities obtained from the two different methods, we confirmed the spin degeneracy $s = 2$ in the 6-uc-thick film without any signature of Zeeman splitting, which is the same as that in the 2.5-uc-thick film (carrier densities extracted from the Landau plot and the Hall curve are $s \times 3.8 \times 10^{12}$ $cm^{-2}$ and $7.1 \times 10^{12}$ $cm^{-2}$). Such absence of Zeeman splitting indicates that the energy gap is smaller than thermal broadening, $(g\mu_B B - \Gamma)/2 < k_B T$ [9], where $\mu_B$ and $k_B$ stand for Bohr magneton and Boltzmann constant, respectively, $\Gamma$ ($= \pi k_B T_D$) stands for the disorder broadening of LLs, and $T_D$ represents the Dingle temperature of ~ 10 K (Fig. S8 to S10). At magnetic fields higher than 50 T, the absence of an observable Zeeman splitting suggests that the effective $g$-factor of $Bi_2O_2Se$ is smaller than 1.28. This is significantly different from other 2D semiconductors such as black phosphorus, InSe and transition metal dichalcogenides [7,9-11] (Table S2).

By applying FFT and inverse FFT analyses of quantum oscillations in the 1-uc-thick device (Fig. S11), we conducted the Landau fan diagram of the two splitting peaks in the FFT spectrum. As shown in Fig. 5g, the carrier densities extracted from the slopes of Landau fans are $s \times 2.1 \times 10^{12}$ cm$^{-2}$ and $s \times 3.5 \times 10^{12}$ cm$^{-2}$. By comparing the carrier density extracted from the Hall curve, $6.0 \times 10^{12}$ cm$^{-2}$, we found that the spin degeneracies of two splitting peaks are $s = 1$, and thus the splitting in the FFT spectrum arises from the global Rashba splitting. Therefore, the $v = 5$ quantum Hall plateau in the 1-uc-thick film is contributed by the inner ($v = 2$) and outer ($v = 3$) Rashba pockets (Fig. S11). The global Rashba parameter exhibits a large value of $\alpha_R^0 \approx \frac{\hbar^2}{m^*}\sqrt{\frac{\pi}{2}\frac{\Delta n}{\sqrt{n}}} \approx 440$ meV·Å, where $\hbar$ is reduced Planck constant, $m^*$ is the effective mass, $\Delta n$ is the difference in densities between majority and minority spin carriers [30]. The analysis of quantum oscillations clearly reveals that the various quantum Hall states appearing in the Bi$_2$O$_2$Se films are related to different spin degeneracies. Note that epitaxial 1-uc-thick Bi$_2$O$_2$Se has one of the strongest global Rashba splitting among known Rashba systems as summarized in Table S3 [14,28,30,41-46], thus can serve as an ideal platform for studying Rashba-related physics such as intrinsic spin Hall effect [47,48], spin interference [49], spin galvanic effect [50], and spin-orbital torque [51,52].

**Theoretical model for the origin of unique QHE**

We adopt a Rashba bilayer model based on *ab initio* band structures of Bi$_2$O$_2$Se thin films to understand the absence of odd-integer quantum Hall plateaus in 2D Bi$_2$O$_2$Se (Fig. S12 to S16 and Ref. 53). An $n$-uc-thick Bi$_2$O$_2$Se film has $4n$ Rashba layers (Fig. 6a). Note that Rashba monolayers couple strongly inside a [Bi$_2$O$_2$]$^{2+}$ layer but weakly between neighboring [Bi$_2$O$_2$]$^{2+}$ layers. The interlayer coupling alternates along the $z$ direction like a Su–Schrieffer–Heeger chain [54], as shown in Fig. 6a. Because of the considerable difference between intra- and inter-layer coupling, we can further downfold the low energy states to $2n$ Rashba layers. In such an effective model, neighboring Rashba layers exhibit opposite Rasbha coefficients ($\pm\alpha$) but uniform interlayer coupling ($t$). Note that the Rashba parameter $\alpha$ originates from the hidden Rashba monolayer in each [Bi$_2$O$_2$]$^{2+}$ layers, and is different from the global Rashba parameter $\alpha_R^0$ due to the interface-driven inversion-breaking. As we will show in the following, the hidden Rashba effect will dramatically suppress the Zeeman splitting in the QHE.

The effective Hamiltonian for a *n*-uc-thick film is,

$$H = \frac{p^2}{2m^*}\delta_{l,l\prime} + (-1)^l \alpha(k_y \sigma_x - k_x \sigma_y)/\hbar - t\delta_{l,l\pm 1}\delta_{\sigma\sigma\prime}. \quad (1)$$

Where $m^*$ is the effective mass, $p^2 = p_x^2 + p_y^2$ are the in-plane momentum, $\sigma_{x,y}$ are Pauli matrices which denote the spin degree of freedom, *l*=1,2,3,...2*n* is the Rashba layer index and $(-1)^l \alpha$ represents the alternating Rashba layers. This model respects both inversion symmetry and time-reversal symmetry and thus gives doubly degenerate bands. The lowest band corresponds to the Fermi surface observed in our experiments. We carried out *ab initio* calculations on Bi$_2$O$_2$Se films of different thicknesses and extracted parameters for Eq. (1), $m^* = 0.14\, m_0$, $t = 0.2$ eV and $\alpha = 1.45$ eV · Å.

The magnetic field (*B*) is treated in the Landau gauge by considering the replacement $p \rightarrow p + eA$ (for electrons) with $A = (0, Bx, 0)$. For multiple Rashba layers, the coupling matrix between $n$ and $m$ LLs reads,

$$H_{nm}^{l,l\prime} = \hbar\omega_c\left(n + \frac{1}{2}\right)\delta_{nm}\delta_{ll\prime} - t\delta_{nm}\delta_{l,l\prime\pm 1} + \frac{g_0 \mu_B B \sigma_z}{2}\delta_{nm} + \frac{(-1)^l\sqrt{2}\alpha}{l_B}\left(\delta_{n,m+1}\sigma^+ + \delta_{n,m-1}\sigma^-\right). \quad (2)$$

Here, $\omega_c = eB/m^*$, $l_B = \sqrt{\hbar/eB}$ are the cyclotron frequency and magnetic length, respectively, $g_0 \approx 2$ is the bare Landé *g*-factor, and $\sigma^\pm = (\sigma_x \pm i\sigma_y)/2$. The Landau-quantized momentum operators renormalize the Rashba interaction such that it couples different LLs and thus competes with the bare Zeeman splitting [third term in Eq. (2)], due to the ladder structure of momentum operators [the fourth term in Eq. (2)]. The induced splitting between $(n, \sigma)$ and $(n, \sigma\prime)$ LLs represents the effective Zeeman energy ($E_Z$). One can verify $E_Z = g_0 \mu_B B$ at $\alpha = 0$ and find that $E_Z$ decreases from $g_0 \mu_B B$ to zero and further to a negative value with increasing $\alpha$.

The effective *g*-factors $g_{eff} = |\Delta\varepsilon/\mu_B B|$ of Bi$_2$O$_2$Se films will be dramatically suppressed in the scenario with spin-degenerated bands. Figure 6b shows the $g_{eff}$ in dependence of $\alpha$ for the 6-uc-thick epitaxial Bi$_2$O$_2$Se films, whose band structure is plotted in Fig. 6c. For the *n*-th LL, $g_{eff}$ first decreases from $g_0$ to zero and further increases with increasing Rashba $\alpha$. From the calculated

Landau fan diagram (Fig. S20), increasing $\alpha$ pushes the upper Zeeman split LL down toward the lower LL and eventually switch their order, leading to $g_{eff}$ trend observed. At the material parameter region of $\alpha$, the small $g_{eff}$ satisfies the experimental requirement for $n \leq 4$ ($g_{eff} <$ 1.28) to suppress the odd-integer Hall plateaus when the films exhibit negligible global Rashba effect. Figure 6d and 6e exhibits the $g_{eff}$ and band structure of the 2.5-uc-thick film, which exhibits weak global Rashba effect. Similar to that in the 6-uc-thick film, the small $g_{eff}$ in the 2.5-uc-thick film also satisfies the experimental requirement for $n \leq 4$ ($g_{eff} < 1.28$) to suppress the odd-integer Hall plateaus.

In the 1-uc-thick Janus $Bi_2O_2Se$ film, the Rashba bilayer scenario breaks down if there exists strong global Rashba effect (Fig. S20, S21 and S22). The origin of thickness-dependent global Rashba effect is summarized in Section 11 in SI. For the 1-uc-thick film, we introduce a symmetry breaking term $\Delta \tau_z$ in Eq. (1), where $\tau_z$ is the Pauli matrix for the layer degree of freedom and $\Delta =$ 78 meV, to reproduce the global Rashba splitting observed in experiments. As shown in Fig. 6f, $g_{eff}$ is still suppressed in the small $\alpha$ region, soon passes the critical point and increases to a large value for the realistic $\alpha$ (the vertical dashed line) in the film with giant global Rashba effect (Fig. 6g). Consequently, we obtain quite large $g_{eff}$ in calculations (Fig. 6f, and S22), rationalizing the appearance of both even and odd plateaus in the quantum Hall regime.

**Conclusions**

In summary, we demonstrated the experimental observation of the absence of odd-integer quantum Hall states originating from the unique degenerated Rashba bilayer structure and hidden Rashba effect in few-layer 2D $Bi_2O_2Se$ films. The quantum Hall states in layered 2D $Bi_2O_2Se$ can be selectively controlled via SOC engineering at atomic level. In the 1-uc-thick Janus $Bi_2O_2Se$ film, inversion-symmetry breaking brings out a giant global Rashba splitting of $\alpha_R^0 \approx 440$ meV·Å, which is one of the largest values among the known 2D semiconducting Rashba systems. The giant global Rashba effect changes the degeneracy of electrons and results in the coexistence of both odd- and even- integer quantum Hall states. Thus, the high mobility, as well as the strong and tunable SOC, make $Bi_2O_2Se$ a promising candidate for realizing novel SOC-related phenomena and potential spintronic applications, such as tunable spin Hall effect, spin galvanic effect, nonlinear physics and spin-orbital torque.

**Acknowledgments:** We acknowledge Molecular Materials and Nanofabrication Laboratory (MMNL) in the College of Chemistry at Peking University for the use of instruments. This work was supported by the National Key Research and Development Program of China (2022YFA1204900 (H.P.), 2021YFA1202901 (J.H. and C.T.)), the National Natural Science Foundation of China (21920102004 (H.P.), 92365203 (H.Y.), 92164205 (J.W. and C.T.), 52021006 (H.P.), 22205011 (C.T.), 52072168 (H.Y.), 51861145201 (H.Y.), 21733001 (H.P.), 22105009 (J.W.) and 52302180 (J.H.)), Beijing National Laboratory for Molecular Sciences (BNLMS-CXTD-202001 (H.P.)), and the Tencent Foundation (The XPLORER PRIZE (H.P.)). B.Y. acknowledges the financial support by the European Research Council (ERC Consolidator Grant "NonlinearTopo", No. 815869) and the Israel Science Foundation (ISF: 2932/21, 2974/23). J.W. acknowledges the support from the Boya Postdoctoral Fellowship. D. K. is supported by the Abrahams Fellowship of the Center for Materials Theory, Rutgers University and the Zuckerman STEM fellowship.

**Author contributions:**

H.P., H.Y. and B.Y. conceived the original idea for the project. X.Z., C.T., Y. L. and X.C synthesized the materials. J.W., X.Z., X.C. and Y.Z. fabricated the devices. X.Z. X.G., R.Z. and P. G. carried out STEM measurements. J.W. and J.H. carried out static magnetic field transport measurements. Pulsed magnetic field transport measurements were performed by J.W., J.H. and C.T. with the help from J.Z., G.J., H.Z. and Z.Z.. D.K., A.S. and B.Y. carried out theoretical calculations. J.W., J.H., X.Z., A.S., H.P., B.Y., and H.Y. wrote the manuscript. All authors contributed to the scientific planning and discussions.

**Competing interests**

Authors declare that they have no competing interests.

**Figures**

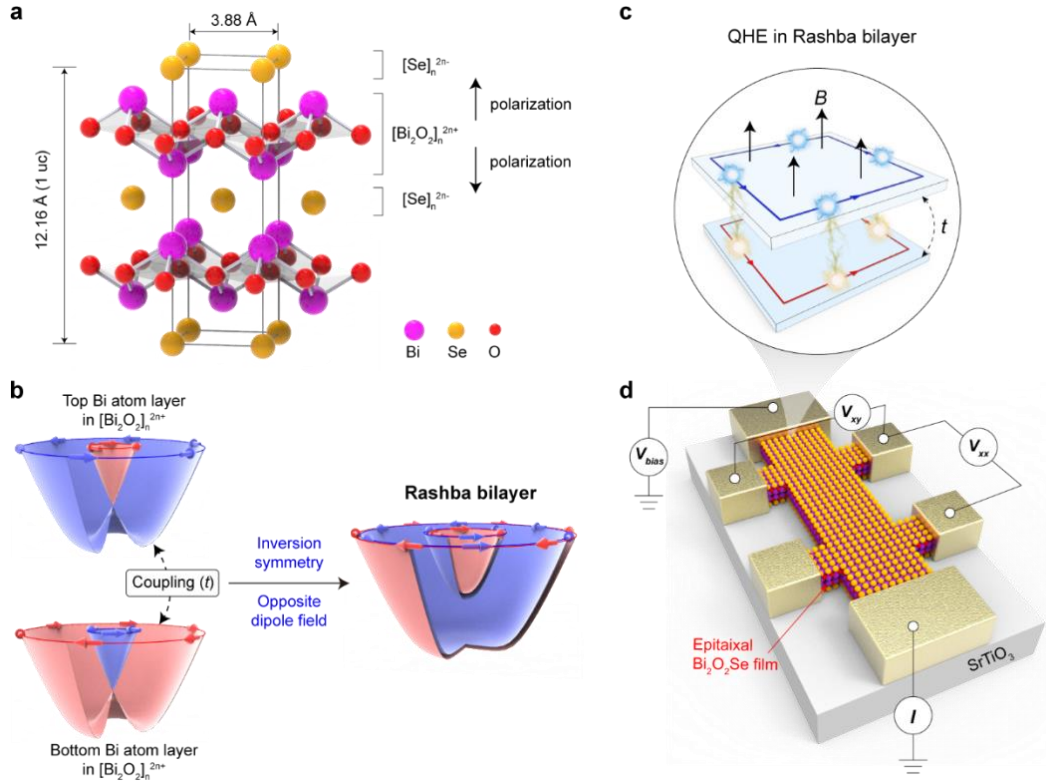

**Fig. 1. Rashba bilayer structure in Bi$_2$O$_2$Se films.** (**a**) Layered crystal structure of 1 unit cell Bi$_2$O$_2$Se with tetragonal [Bi$_2$O$_2$]$_n^{2n+}$ layers and [Se]$_n^{2n-}$ layers alternately stacked along the *c* axis. (**b**) In [Bi$_2$O$_2$]$_n^{2n+}$ layers, each Bi monolayer feels a strong Rashba effect while neighboring Bi monolayer exhibits opposite sign in Rashba SOC, resulting in a hidden Rashba effect as well as unique QHE. The bands in different colors represent bands with different spin textures. The arrows stand for the directions of spin. (**c**) In the schematic view of QHE in Rashba bilayer, electrons in two Rashba layers have different spin textures (colors). (**d**) Schematic view of the Hall bar device for an epitaxial Bi$_2$O$_2$Se film.

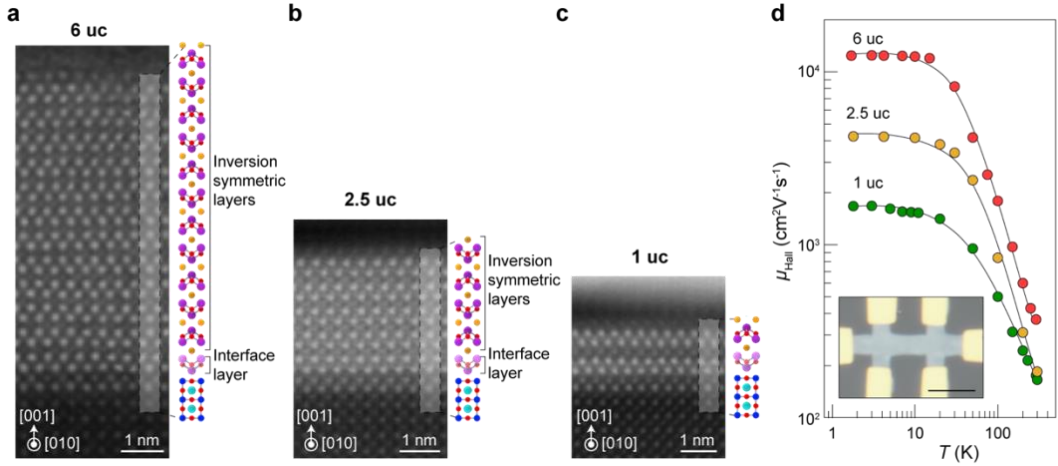

**Fig. 2. STEM images and electrical characterization of epitaxial $Bi_2O_2Se$ films on $SrTiO_3$.** (**a-c**) Cross-sectional HAADF-STEM images and crystal structures of $Bi_2O_2Se$ films with different thicknesses epitaxially grown $SrTiO_3$. The inversion-asymmetric layers in the crystal structures are plotted in a lighter color. (**d**) $\mu_{Hall}$ as a function of temperature for non-encapsulated $Bi_2O_2Se$ films. The Hall mobility of epitaxial $Bi_2O_2Se$ films reach maximums of $\mu_{Hall}$ = 12435, 4230, 1677 $cm^2\ V^{-1}\ s^{-1}$ below 2 K for film thicknesses of 6, 2.5, and 1 uc, respectively. The error is smaller than the data point and the error comes from the fitting of Hall curves. Inset shows the optical image of a typical Hall bar device for an epitaxial $Bi_2O_2Se$ film. Scale bar: 5 μm. The black lines are guides to the eyes.

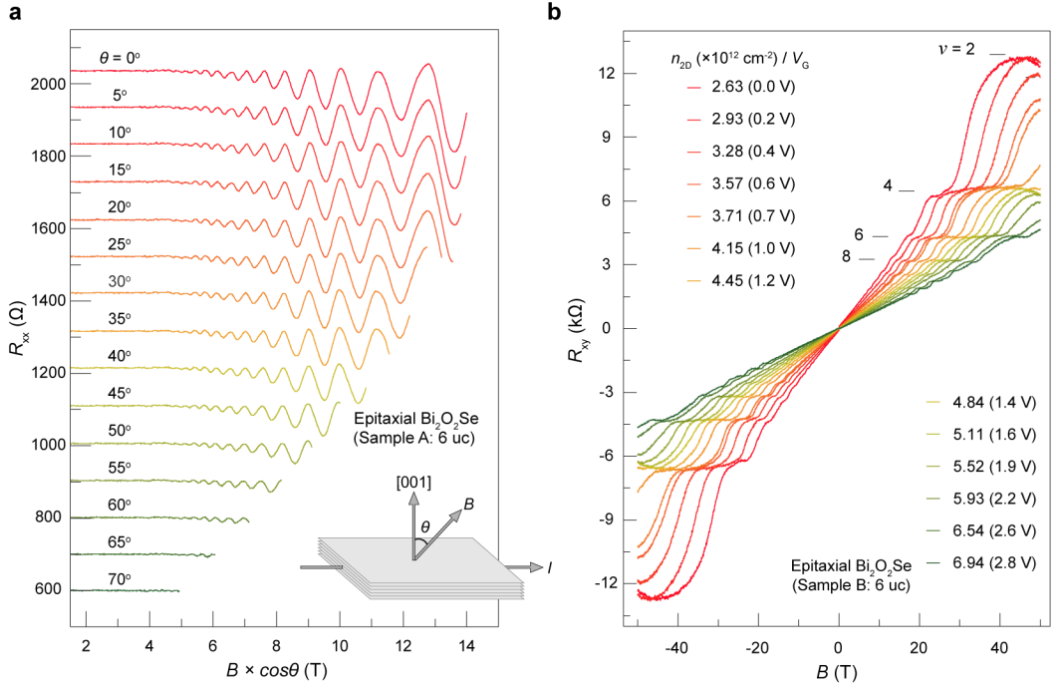

**Fig. 3. Absence of odd-integer QHE in the epitaxial Bi$_2$O$_2$Se films.** (**a**) Angle-dependent magnetoresistance as a function of effective magnetic fields ($B_\perp = B \times \cos\theta$). Inset, $\theta$ is the angle between the magnetic field and the normal direction of the sample plane. The curves are shifted by 100 Ω vertically. The positions of the SdH oscillation extrema correspond to unchanged $B_\perp$ under different $\theta$. (**b**) Hall resistance as a function of the magnetic field measured at varying electron densities in the 6-uc-thick device. The sample was immersed in liquid helium at 4.2 K when applying the pulsed magnetic field up to 50 T. The horizontal lines indicate the resistance related to different even-integer filling factors.

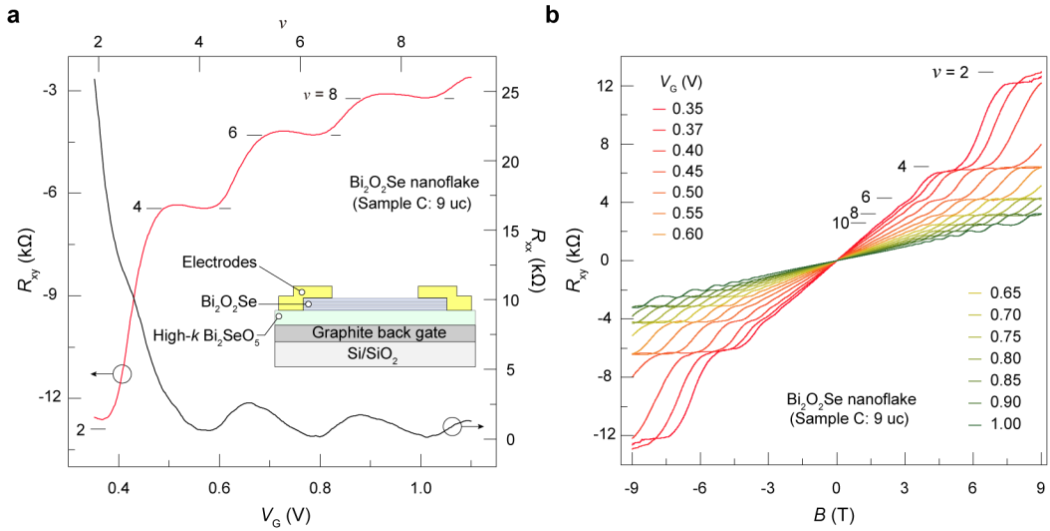

**Fig. 4. Even-integer QHE in Bi$_2$O$_2$Se nanoflakes.** (**a**) Hall resistance and longitudinal resistance as a function of gate voltage at a certain magnetic field $B = 9$ T. Quantum Hall plateaus are clearly observed with filling factors ranging from $v = 2$ to $v = 8$ in a 9-uc-thick free-standing Bi$_2$O$_2$Se nanoflake. Inset: schematic of a back-gate Bi$_2$O$_2$Se Hall bar device on high-$\kappa$ Bi$_2$SeO$_5$ nanoflake. (**b**) Hall resistance as a function of magnetic field. Quantum Hall plateaus are clearly observed with $v$ ranging from 2 to 10.

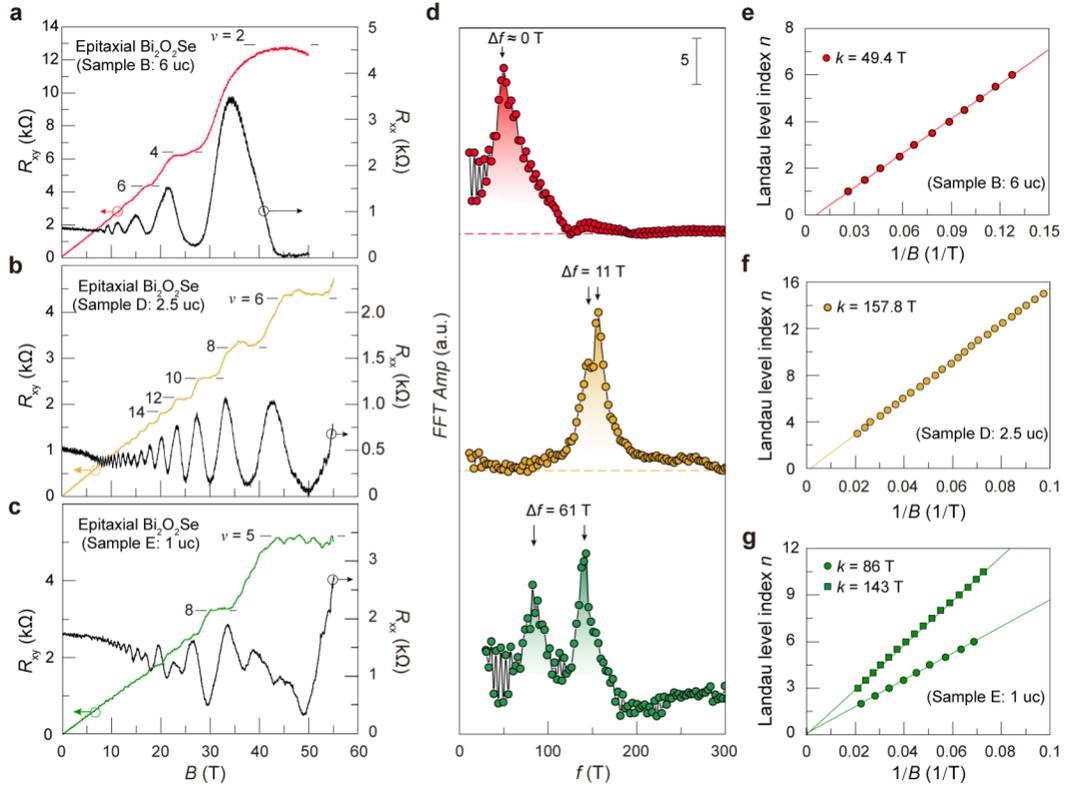

**Fig. 5. Thickness-dependent QHE and quantum oscillations in epitaxial $Bi_2O_2Se$ films.** (**a**) Even-integer QHE in the 6-uc-thick sample at 4.2 K. $R_{xx}$ decreases to 0 when $B > 42$ T. (**b**) Even-integer QHE in the 2.5-uc-thick sample at 4.2 K. The plateau of $R_{xy}$ at $v = 6$ corresponds to the position where $R_{xx}$ decreases to almost 0. (**c**) QHE in the 1-uc-thick sample at 4.2 K. $R_{xy}$ shows quantized plateaus with both even and odd filling factors. (**d**) FFT analysis of SdH oscillations in 6- (red), 2.5- (yellow) and 1-uc-thick (green) devices. (**e**) Landau plots for the SdH oscillations in (**a**). To construct the Landau fan diagram, the resistance $R_{xx}$ and $R_{xy}$ was firstly converted into conductance with the anisotropic Hall tensor, and then valleys of conductance are assigned with integers while the peaks are assigned with half integers. (**f**) Landau plots for the SdH oscillations in the 2.5-uc-thick device. (**g**) Landau plots in the 1-uc-thick device. Two Landau fans originate from the two splitting peaks in the lowest panel in (**d**).

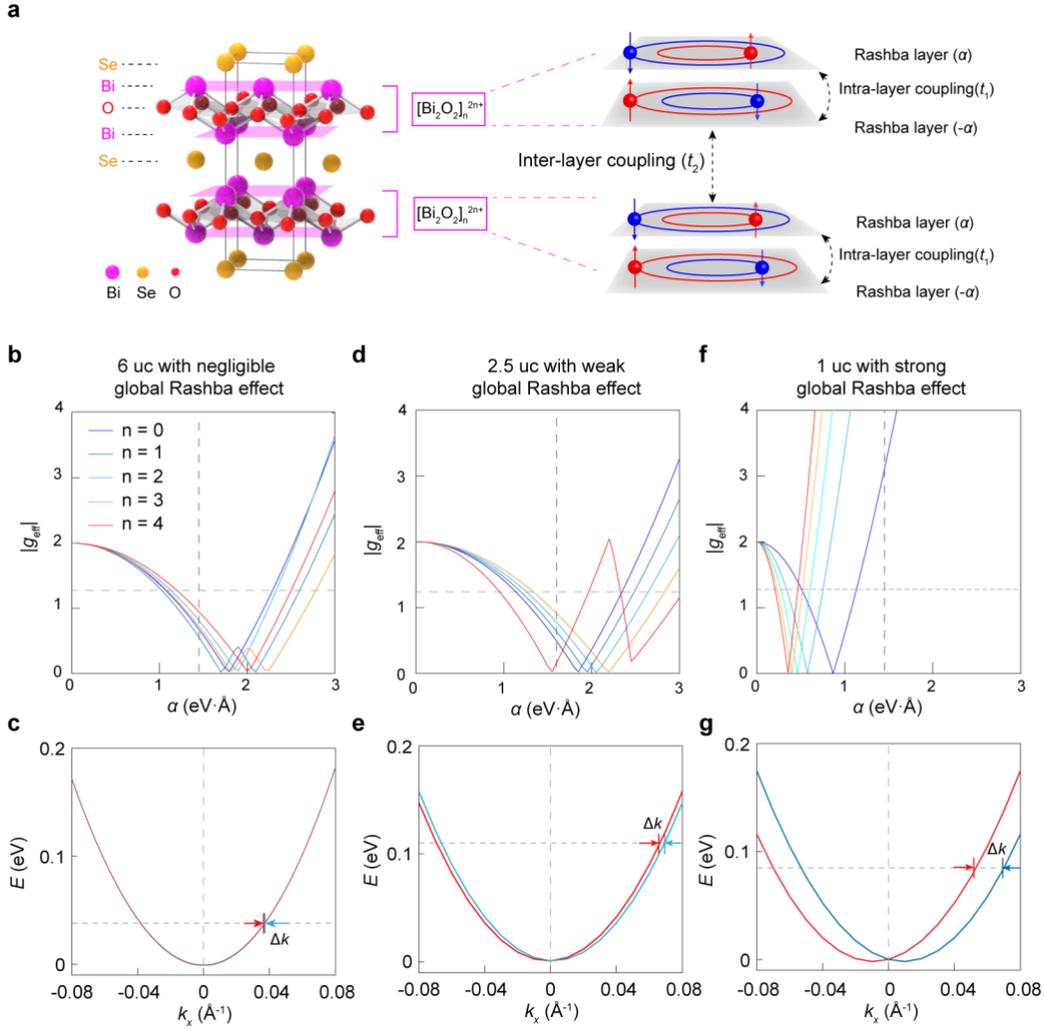

**Fig. 6. The calculated effective *g*-factors and band structures for Bi$_2$O$_2$Se thin films**. (**a**) Unit cell of bulk Bi$_2$O$_2$Se composed Bi-O layers with alternating spin orbital coupling $(\alpha, -\alpha, \alpha, -\alpha)$. Each bilayer is inversion-symmetric. The intra-layer coupling is denoted by $t_1$ while the inter-layer coupling is $t_2$. (**b**), (**d**), (**f**) The $g_{eff}$ dependence on Rashba parameter $\alpha$ for different films. The vertical dashed line indicates the $\alpha$ value in materials. The horizontal dashed line indicates the upper limit of $g_{eff} = 1.28$ to suppress the Zeeman splitting. (**c**), (**e**), (**g**) Lowest energy band structures for epitaxial Bi$_2$O$_2$Se films with different thicknesses, which were derived from the experiments and *ab initio* calculations. The separation between two arrows represents the difference of wave vector ($\Delta k$) due to Rashba splitting, which increases when reducing the film thickness. Lines in different colors represent bands with different spin directions.

**Materials and Methods**

$Bi_2O_2Se$ film growth and characterization

The $Bi_2O_2Se$ films were grown in a homemade oxide molecular beam epitaxy (OMBE) system with base pressure better than $5\times10^{-10}$ mbar and the detailed growth condition is the same as that in our previous work [33]. AFM images were performed on a Bruker Nanoscope system (Dimension Icon). The X-ray diffraction (XRD) measurements were performed on a Rigaku SmartLab (9 kW) X-ray diffractometer with a Ge (220) × 2 crystal monochromator. The STEM measurements were performed on FEI Titan Cubed Themis G2 at 300 kV.

Fabrication of Hall devices

The fabrication process of Hall device of MBE $Bi_2O_2Se$ films is as follows: MBE-grown $Bi_2O_2Se$ films were etched into discrete $40 \times 20$ μm$^2$ rectangles by maskless laser direct writing (MLDW) and wet chemical etching (WCE) method [33,55]. The etchant is $H_2SO_4$: $H_2O_2$: $H_2O$=1: 2: 4. Then Hall bar patterns were fabricated by standard electron beam lithography (EBL, FEI Quanta 250FEG) and etching process. Then, standard EBL technique was used to fabricate six-terminal Hall-bar metal contacts. Pd/Au contact electrodes (7 nm/45 nm) were finally deposited by E-beam evaporation (DE400). In order to avoid the influence from the substrate steps, all the devices were fabricated on individual terraces with the device channels parallel to the steps. The width of Hall bar ranges from 1.5 to 3 microns, the length of the channel ranges from 5 to 10 microns.

The fabrication process of Hall device of free-standing $Bi_2O_2Se$ nanoflake is as follows: $Bi_2O_2Se$ nanoflake was transferred on pre-exfoliated high-$\kappa$ van der Waals layered $Bi_2SeO_5$ nanoflake with graphite back gate. Then, EBL and E-beam evaporation were used to prepare multiple metal contacts for six-terminal Hall-bar structures [40].

Electrical transport measurements

Electrical transport measurements were carried out in the following systems: an integrated cryofree static superconducting magnet system (14-T TeslatronPT from Oxford Instruments, or 9-T AttoDry2100 from Attocube Systems), and the pulsed high magnetic field facility (55 T) at Wuhan National High Magnetic Field Center. An *in-situ* annealing process was necessary before the electrical transport measurements. Namely, the sample was mounted into the chamber, and then pumped to a high vacuum better than $10^{-3}$ mbar and heated to 385 K. The sample was kept at above conditions for about 1 hour, and the resistance was decreased rapidly to a constant value. Subsequently, the sample was cooled down to the base temperature of the cryogenic systems. The gate voltage was applied by Keithley 2400 on the $Bi_2O_2Se$ devices with STO (001) or $Bi_2SeO_5$ as the gate dielectric. At static magnetic fields, the electrical transport signals were detected by using multiple lock-in amplifiers (Stanford Research Systems, SR830) with an AC excitation current ($I$) below 1 μA at a frequency between 10 and 20 Hz. Under pulsed high magnetic fields, the $Bi_2O_2Se$ devices were immersed into liquid helium to avoid eddy-current heating, and the electric resistance was measured by using a DC technique with a current of 2.5 μA. Four pulsed shots for the magnetoresistance measurements at different conditions (+$I$+$B$, +$I$–$B$, –$I$+$B$, –$I$–$B$) were performed to obtain pure QHE signals. Results from different systems and different samples are reproducible and consistent. As applying a large gate voltage will change the inversion symmetry and the global Rashba parameters, all the quantum transport measurements were performed without applying a gate voltage in figure 5.

**Data and materials availability**

All data needed to evaluate the conclusions in this paper are present in the paper or the supplementary materials. Source data are provided with this paper.

**Methods-only references**